\documentstyle[twoside,fleqn,epsfig]{actaps}

\begin{document}

\newcommand{\beq}{\begin{equation}}\newcommand{\eeq}{\end{equation}}
\newcommand{\barr}{\begin{eqnarray}}\newcommand{\earr}{\end{eqnarray}}
\newcommand{\andy}[1]{ }
\def\ask{\marginpar{?? ask:  \hfill}}  \def\fin{\marginpar{fill in ...
\hfill}}
\def\note{\marginpar{note \hfill}}  \def\check{\marginpar{check \hfill}}
\def\discuss{\marginpar{discuss \hfill}}
\def\hh{\widehat}
\def\wtilde{\widetilde}
\newcommand{\bm}[1]{\mbox{\boldmath $#1$}}
\newcommand{\bmsub}[1]{\mbox{\boldmath\scriptsize $#1$}}
\newcommand{\bmh}[1]{\mbox{\boldmath $\hat{#1}$}}

\headings{1}{8}
\def\authorlist{S Pascazio, P Facchi}
\def\shorttitle{Modifying lifetimes}

\title{\uppercase{Modifying the lifetime of an unstable system by an intense electromagnetic field}}

\author{S. Pascazio\email{saverio.pascazio@ba.infn.it},
P. Facchi\email{paolo.facchi@ba.infn.it}}
{ Dipartimento di Fisica, Universit\`a di Bari \\
        Istituto Nazionale di Fisica Nucleare, Sezione di Bari \\
 I-70126  Bari, Italy
}

\day{30 April 1999}  

\abstract{%
We study the temporal behavior of a three-level system (such as an
atom or a molecule), initially prepared in an excited state, bathed
in a laser field tuned at the transition frequency of the other
level. We analyze the dependence of the lifetime of the initial
state on the intensity of the laser field. The phenomenon we
discuss is related to both electromagnetic induced transparency and
quantum Zeno effect. }

\medskip

\pacs{42.50.Hz; 42.50.Vk; 03.65.Bz}

\section{Introduction}

The temporal behavior of quantum mechanical systems, being governed
by unitary operators \cite{Dirac}, displays some subtle features at
short \cite{shortt} and long times \cite{longtt}. In order to
discuss the evolution of genuine unstable systems one usually makes
use of the Weisskopf-Wigner approximation \cite{seminal}, which
ascribes the main properties of the decay law to a pole located
near the real axis of the complex energy plane. This yields the
Fermi ``Golden Rule" \cite{Fermigold}. These features of the
quantum evolution are well known and discussed in textbooks of
quantum mechanics \cite{Sakurai} and quantum field theory
\cite{Brown}. For a recent review, see \cite{temprevi}. In this
paper we shall investigate the possibility that the lifetime of an
unstable quantum system can be modified by the presence of a very
intense electromagnetic field. We shall look at the temporal
behavior of a three-level system (such as an atom or a molecule),
where level \#1 is the ground state and levels \#2, \#3 are two
excited states. The system is initially prepared in level \#2 and
if it follows its natural evolution, it will decay to level \#1.
The decay will be (approximately) exponential and characterized by
a certain lifetime, that can be calculated from the Fermi Golden
Rule. But what happens if one shines on the system an intense laser
field, tuned at the transition frequency 3-1? This problem was
investigated in Ref.\ \cite{MPS}, in the context of the so-called
quantum Zeno effect \cite{QZE}. It was found that the lifetime of
the initial state depends on the intensity of the laser field. In
the limit of extremely intense field, the decay should be
considerably slowed down. The aim of this paper is to study this
effect in more detail and discuss a new phenomenon: we shall see
that for physically sensible values of the laser field, the decay
can be {\em enhanced}, rather than hindered. This can be viewed as
an ``inverse" quantum Zeno effect. The whole problem is related to
electromagnetic induced transparency \cite{induced}.

\section{Preliminaries and definitions }
 \label{sec-preldef}
 \andy{preldef}

We start from the Hamiltonian \cite{MPS}:
\andy{ondarothamdip6}
\barr
H &=& H_{0}+H_{\rm int}\nonumber\\
&=& \omega_0|2\rangle\langle 2|+\Omega_0|3\rangle\langle 3|
+\sum_{\bmsub k,\lambda}\omega_k a^\dagger_{\bmsub{k}\lambda}
a_{\bmsub{k}\lambda}
+\sum_{\bmsub k,\lambda}\left(\phi_{\bmsub k\lambda}
a_{\bmsub k\lambda}^\dagger|1\rangle\langle2|
+\phi_{\bmsub k\lambda}^* a_{\bmsub
k\lambda}|2\rangle\langle1|\right)\nonumber\\ & &+\sum_{\bmsub
k,\lambda}\left(\Phi_{\bmsub k\lambda}
a_{\bmsub k\lambda}^\dagger|1\rangle\langle3|
+\Phi_{\bmsub k\lambda}^* a_{\bmsub
k\lambda}|3\rangle\langle1|\right),
\label{eq:ondarothamdip6}
\earr
where the first two terms are the free Hamiltonian of the 3-level
system (whose states $|i\rangle$ $(i=1,2,3)$ have energies $E_1=0$,
$\omega_0=E_2-E_1>0$, $\Omega_0=E_3-E_1>0$), the third term is the
free Hamiltonian of the EM field and the last two terms describe
the $1\leftrightarrow2$ and $1\leftrightarrow3$ transitions in the
rotating wave approximation, respectively. The states $|2\rangle$
and $|3\rangle$ are chosen so that no transition between them is
possible (e.g., because of selection rules). The matrix elements of
the interaction Hamiltonian read
\andy{intel}
\barr
\phi_{\bmsub k\lambda}&=&\frac{e}{\sqrt{2\epsilon_0 V\omega}}
\int d^3x\;e^{-i\bmsub k\cdot\bmsub x}\bm\epsilon_{\bmsub k\lambda}^*\cdot \bm
j_{12}(\bm x) ,
\nonumber \\
\Phi_{\bmsub k\lambda}&=&\frac{e}{\sqrt{2\epsilon_0 V\omega}}
\int d^3x\;e^{-i\bmsub k\cdot\bmsub x}\bm\epsilon_{\bmsub k\lambda}^*\cdot \bm
j_{13}(\bm x),
\label{eq:intel}
\earr
where $-e$ is the electron charge, $\epsilon_0$ the vacuum
permittivity, $V$ the volume of the box, $\omega=|\bm k|$,
$\bm\epsilon_{\bmsub k\lambda}$ the photon polarization and $\bm
j_{\em fi}$ the transition current of the radiant system. For
example, in the case of an electron moving in an external field, we
have $\bm j_{\em fi}=\psi_{\em f}^\dagger \bm\alpha\psi_{\em i}$
where $\psi_{\em i}$ and $\psi_{\em f}$ are the electron
wavefunctions of the initial and final states, respectively, and
the components of $\bm\alpha$ are the usual Dirac matrices. For the
sake of generality we are using relativistic matrix elements
(although our analysis can be performed with nonrelativistic ones
$\bm j_{\em fi}=\psi_{\em f}^*\bm p\psi_{\em i}/m_e$, where $\bm
p/m_e$ is the electron velocity).

We shall concentrate our attention on a 3-level system bathed in a continuous
laser beam, whose photons have momentum
$\bm k_0$ ($|\bm k_0|=\Omega_0$) and polarization $\lambda_0$.
We shall also assume, throughout this paper, that
\andy{noint}
\beq
\phi_{\bmsub k_0\lambda_0}=0,
\label{eq:noint}
\eeq
i.e., the laser does not interact with state $|2\rangle$. Also,
since the average number $N_0$ of $\bm k_0$-photons in the total
volume $V$ can be considered very large, we shall perform our
analysis in terms of number (rather than coherent) states of the EM
field \cite{Knight}. In this approximation,
\andy{appros}
\barr
& & \langle1; 0_{\bmsub k\lambda}, N_0|H_{\rm int} |3; 0_{\bmsub
k\lambda},N_0-1\rangle=\sqrt{N_0}
\Phi_{\bmsub k_0\lambda_0}
\nonumber \\
& & \qquad \qquad \gg
\langle1;1_{\bmsub k\lambda}, N_0-1|H_{\rm int}
|3;0_{\bmsub k\lambda},N_0-1\rangle=
\Phi_{\bmsub k\lambda},
\label{eq:appros}
\earr
$\forall(\bm k,\lambda)\neq(\bm k_0,\lambda_0)$. In the above
equation and henceforth, the vector $|i;n_{\bmsub k \lambda},
M_0\rangle$ represents an atom or a molecule in state $|i\rangle$,
with $n_{\bmsub k \lambda}$ $(\bm k,\lambda)$-photons and $M_0$
laser photons.

In the above approximation, the Hamiltonian (\ref{eq:ondarothamdip6})
can be replaced by
\andy{ondarothamdip7}
\barr
H \! &\simeq& \! \omega_0|2\rangle\langle
2|+\Omega_0|3\rangle\langle 3| +\sum_{\bmsub k,\lambda}\omega_k
{a}^\dagger_ {\bmsub{k}\lambda} {a}_ {\bmsub{k}\lambda}
+{\sum_{\bmsub k,\lambda}}'\left(\phi_{\bmsub k\lambda} a_{\bmsub
k\lambda}^\dagger|1\rangle\langle2| +\phi_{\bmsub k\lambda}^*
a_{\bmsub k\lambda}|2\rangle\langle1|\right)\nonumber\\ &
&+\left(\Phi_{\bmsub k_0\lambda_0} a_{\bmsub
k_0\lambda_0}^\dagger|1\rangle\langle3| +\Phi_{\bmsub
k_0\lambda_0}^* a_{\bmsub k_0\lambda_0}|3\rangle\langle1|\right),
\label{eq:ondarothamdip7}
\earr
where the prime means that the summation does not include $(\bm
k_0,\lambda_0)$ [due to our hypothesis (\ref{eq:noint})].
The operators
\andy{operatoreN}
\beq
{\cal N}=|2\rangle\langle 2|+{\sum_{\bmsub k,\lambda}}'
{a}^\dagger_{\bmsub{k}\lambda}{a}_{\bmsub{k}\lambda},
\qquad
{\cal N}_0=|3\rangle\langle 3|+{a}^\dagger_{\bmsub{k}_0\lambda_0}
{a}_{\bmsub{k}_0\lambda_0}
\label{eq:operatoreN}
\eeq
satisfy
\beq
[H,{\cal N}]=[H,{\cal N}_0]=[{\cal
N}_0,{\cal N}]=0,
\eeq
which imply the conservation of the total number of photons plus
the atomic excitation (Tamm-Dancoff approximation
\cite{TammDancoff}). The Hilbert space splits therefore into
sectors that are invariant under the action of the Hamiltonian: in
our case, the system evolves in the subspace labelled by the
eigenvalues ${\cal N}=1$, ${\cal N}_0=N_0$ and the analysis can be
restricted to this sector \cite{Knight}.

\section{Temporal evolution}
\label{tempevol}
\andy{tempevol}

The states of the total system in the sector $({\cal N}, {\cal
N}_0)=(1,N_0)$ read
\label{statesdefin}
\beq\label{eq:statesdefin}
|\psi(t)\rangle=x(t)|2;0,N_0\rangle+{\sum_{\bmsub
k,\lambda}}'y_{\bmsub k\lambda}(t)|1;1_{\bmsub
k\lambda},N_0\rangle+{\sum_{\bmsub k,\lambda}}'z_{\bmsub
k\lambda}(t)|3;1_{\bmsub k\lambda},N_0-1\rangle,
\eeq
with the normalization
\andy{normpsi6}
\beq\label{eq:normpsi6}
\langle\psi(t)|\psi(t)\rangle=|x(t)|^2+{\sum_{\bmsub k,\lambda}}'|y_{\bmsub k,
\lambda}(t)|^2+{\sum_{\bmsub k,\lambda}}'|z_{\bmsub k,
\lambda}(t)|^2=1,\qquad \forall t.
\eeq
At time $t=0$ we prepare our system in the state
\andy{condinxy6}
\beq\label{eq:condinxy6}
|\psi(0)\rangle=|2;0,N_0\rangle\quad\Leftrightarrow\quad
x(0)=1,\;y_{\bmsub k\lambda}(0)=0,\;z_{\bmsub k\lambda}(0)=0,
\eeq
which is an eigenstate of the free Hamiltonian
\beq
H_0|\psi(0)\rangle=H_0|2;0,N_0\rangle=\omega_0|2;0,N_0\rangle.
\eeq
We set, without any loss of generality, $E_1+N_0\Omega_0=0$. By
inserting (\ref{eq:statesdefin}) in the time-dependent
Schr\"odinger equation
\beq
i\frac{d}{dt}|\psi(t)\rangle= H|\psi(t)\rangle
\eeq
and Laplace transforming with the initial conditions
(\ref{eq:condinxy6}), one readily obtains
\andy{xs,ys,zs}
\barr
\wtilde x(s)&=&\frac{1}{s+i\omega_0+Q(B,s)},\label{eq:xs}\\
\wtilde y_{\bmsub k\lambda}(s)&=&\frac{-i\phi_{\bmsub k\lambda}(s+i\omega_k)}
{(s+i\omega_k)^2+B^2}\;\wtilde x(s),\label{eq:ys}\\
\wtilde z_{\bmsub k\lambda}(s)&=&-\frac{\sqrt{N_0}\Phi^*_{\bmsub k_0\lambda_0}
\phi_{\bmsub k\lambda}}
{(s+i\omega_k)^2+B^2}\;\wtilde x(s),
\label{eq:zs}
\earr
where the tilde denotes Laplace transform and
\andy{Q(B,s)dipdiscr}
\beq
\label{eq:Q(B,s)dipdiscr}
Q(B,s)=\sum_{\bmsub k,\lambda}|\phi_{\bmsub
k\lambda}|^2\frac{s+i\omega_k}{(s+i\omega_k)^2+B^2},
\qquad
B^2=N_0\,|\Phi_{\bmsub k_0\lambda_0}|^2.
\eeq
$B$ is the intensity of the laser field and can be viewed as the
``strength" of the observation performed by the laser beam on level
\#2 \cite{MPS}.

In the continuum limit ($V\rightarrow\infty$), the matrix elements scale
as follows
\andy{chi}
\beq\label{eq:chi}
\lim_{V\rightarrow\infty}\frac{V\omega^2}{(2\pi)^3}\sum_\lambda\int d \Omega
|\phi_{\bmsub k\lambda}|^2 \equiv g^2\omega_0\chi^2(\omega) ,
\eeq
where $\Omega$ is the solid angle. The (dimensionless) function
$\chi(\omega)$ and the coupling constant $g$ have the following
general properties:
\andy{chiprop,g2}
\barr
\chi^2(\omega) & \propto &
\left\{ \begin{array} {ll}
         \omega^{2j\mp 1} &  \quad \mbox{if $\omega \ll \Lambda$}\\
         \omega^{-\beta} & \quad \mbox{if $\omega \gg \Lambda$}
\end{array}
\right. ,
\label{eq:chiprop}\\
g^2 &=& \alpha (\omega_0 /\Lambda )^{2j+1\mp 1} , \label{eq:g2}
\earr
where $j$ is the total angular momentum of the photon emitted in
the $2\rightarrow 1$ transition, $\mp$ represent electric and
magnetic transitions, respectively, $\beta (> 1)$ is a constant,
$\alpha$ the fine structure constant and $\Lambda$ a natural cutoff
(e.g., of the order of the inverse Bohr radius), which determines
the range of the atomic or molecular form factor. The above
equations are due to general properties of the matrix elements
\cite{BLP,FPinduced}.

In order to scale the quantity $B$, we take the limit of very large cavity,
by keeping the density of $\Omega_0$-photons in the cavity constant:
\andy{limterm}
\beq\label{eq:limterm}
V\rightarrow\infty,\qquad
N_0\rightarrow\infty,\quad\mbox{with}\quad
\frac{N_0}{V}=n_0=\mbox{const} .
\eeq
We obtain from (\ref{eq:Q(B,s)dipdiscr})
\beq
B^2 = n_0 V |\Phi_{\bmsub k_0\lambda_0}|^2 =
(2\pi)^3 n_0 |\varphi_{\lambda_0} (\bm k_0)|^2
\eeq
where $\varphi$ is the scaled matrix element of the 1-3 transition.
As we shall see, in order to affect significantly the lifetime of
level \#2, we shall need a high value of $B$. It is therefore of
interest to consider a 1-3 transition of the dipole type, in which
case the above formula reduces to
\beq
B^2 = 2\pi\alpha\Omega_0|\bm{\epsilon}_{\bmsub{k_0}\lambda_0}^*
\cdot\bm x_{13}|^2 n_0,
\eeq
where $\bm x_{13}$ is the dipole matrix element.

\subsection{Laser off}
\label{laseroff}
\andy{laseroff}

\begin{figure}
\centerline{\epsfig{file=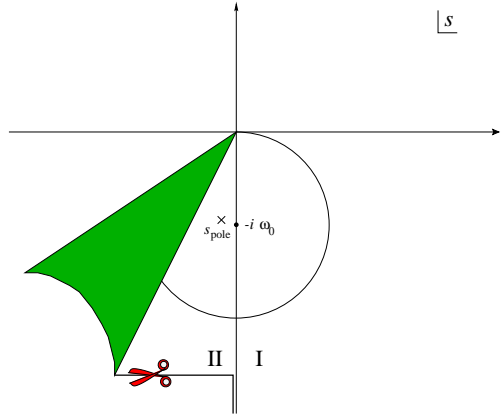,height=5.5cm}}
\caption{%
Fig.\ 3.1. Cut and pole in the $s$-plane ($B=0$). I and II are the
first and second Riemann sheets, respectively. }
\label{fig:fig1}
\end{figure}

Let us first look at the case $B=0$. The laser, tuned at the 1-3
frequency $\Omega_0$, is off and we expect to recover the
well-known physics of a two-level system prepared in an excited
state and coupled to the radiation field \cite{FP1}. In this case,
$Q(0,s)$ is nothing but the self-energy function
\andy{sef}
\beq
Q(s)\equiv Q(0,s)=\sum_{\bmsub k,\lambda}|\phi_{\bmsub
k\lambda}|^2\frac{1}{s+i\omega_k}
\label{eq:sef}
\eeq
and becomes, in the continuum limit,
\andy{Q(s)cont}
\beq\label{eq:Q(s)cont}
Q(s) \equiv g^2 \omega_0 q(s)\equiv
-i g^2 \omega_0 \int_0^\infty d\omega
\frac{\chi^2(\omega)}{\omega-is},
\eeq
where $\chi$ is defined in (\ref{eq:chi}). The function $\wtilde
x(s)$ in Eq.\ (\ref{eq:xs}) (with $B=0$) has a logarithmic branch
cut extending from 0 to $-i\infty$, no singularities on the first
Riemann sheet (physical sheet) and a simple pole on the second
Riemann sheet. The pole equation is
\beq
s+i\omega_0+g^2\omega_0 q_{\rm II}(s)=0,
\eeq
where
\beq
q_{\rm II}(s)=q(s e^{-2\pi i})=q(s)+2 \pi\chi^2 (is)
\eeq
is the determination of $q(s)$ on the second Riemann sheet. We note
that $g^2 q(s)$ is $O(g^2)$, so that the pole can be found
perturbatively: by expanding $q_{\rm II}(s)$ around $-i\omega_0$ we
get a power series, whose radius of convergence is $R_c=\omega_0$
because of the branch point at the origin. The circle of
convergence lies half on the first Riemann sheet and half on the
second sheet (Figure \ref{fig:fig1}).
The pole is well inside the convergence circle,
because $|s_{\rm pole}+i\omega_0|\sim g^2\omega_0\ll R_c$, and
we can write
\beq
s_{\rm pole}= -i\omega_0-g^2\omega_0 q_{\rm II}(-i\omega_0-0^+)+O(g^4)
=-i\omega_0-g^2 \omega_0 q(-i\omega_0+0^+)+O(g^4),
\eeq
because $q_{\rm II}(s)$ is the analytical continuation of $q(s)$
below the branch cut. By using the formula
\beq
\lim_{\varepsilon\rightarrow 0^+}\frac{1}{x\pm i\varepsilon}=
P\frac{1}{x}\mp i\pi\delta(x),
\eeq
one gets from (\ref{eq:Q(s)cont})
\andy{Q(-ieta)}
\barr
q(-i\eta+0^+)&=&-i\int_0^\infty d\omega\,\chi^2(\omega)\frac{1}
{\omega-\eta-i0^+}\nonumber\\ &=&\pi
\chi^2(\eta)\theta(\eta)-iP\int_0^\infty d\omega\,
\chi^2(\omega)\frac{1} {\omega-\eta}
\label{eq:Q(-ieta)}
\earr
and by setting
\andy{spole}
\beq\label{eq:spole}
s_{\rm pole}=-i\omega_0+i\Delta E-\frac{\gamma}{2},
\eeq
one gets
\andy{Fgr}
\beq
\gamma=2\pi g^2\omega_0\chi^2(\omega_0)+O(g^4),
\qquad
\Delta E=g^2\omega_0 P \int_0^\infty\frac{\chi^2(\omega)}{\omega-\omega_0}
+O(g^4),
\label{eq:Fgr}
\eeq
which are the Fermi Golden Rule and the second order correction to
the energy $\omega_0$ of level \#2.

\subsection{Laser on}
\andy{laseron}

We now turn our attention to the situation with the laser switched
on, $B\neq0$. The self energy function $Q(B,s)$ in
(\ref{eq:Q(B,s)dipdiscr}) depends on the value of $B$ and can be
written in terms of the self energy function $Q(s)$ in absence of
laser field [Eq.\ (\ref{eq:sef})], by making use of the following
remarkable property:
\andy{propnotev}
\barr
Q(B,s)&=&\frac{1}{2}\sum_{\bmsub k,\lambda}|\phi_{\bmsub
k\lambda}|^2\left(\frac{1}{s+i\omega_k+iB}+\frac{1}{s+i\omega_k-iB}\right)
\nonumber \\
&=& \frac{1}{2}\left[Q(s+iB)+Q(s-iB)\right].
\label{eq:propnotev}
\earr
Notice, incidentally, that in the continuum limit ($V\to\infty$), due to the above formula, $Q(B,s)$ scales
like $Q(s)$.
The position of the pole $s_{\rm pole}$
(and as a consequence the lifetime $\tau_{\rm
E}\equiv\gamma^{-1}=-1/2\Re e s_{\rm pole}$) depends on the value of $B$.
There are now two branch cuts in the complex $s$
plane, due to the two terms in (\ref{eq:propnotev}). They lie over
the imaginary axis, along $(-i\infty,-iB]$ and
$(-i\infty,+iB]$.

\begin{figure}
\centerline{\epsfig{file=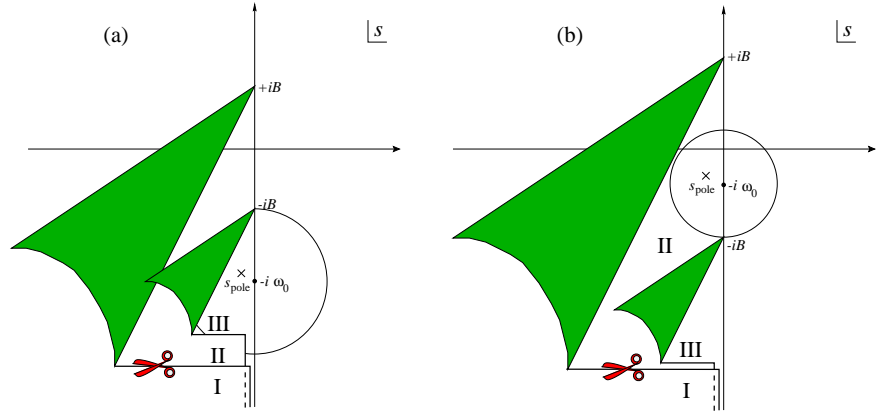,height=5.5cm}}
\caption{%
Fig.\ 3.2. Cuts and pole in the $s$-plane ($B \neq 0$) and
convergence circle for the expansion of $Q(B,s)$ around
$s=-i\omega_0$. I , II and III are the first, second and third
Riemann sheets, respectively. (a) $B<\omega_0$. (b) $B>\omega_0$.}
\label{fig:tagli}
\end{figure}

The pole satisfies the equation
\beq
s + i\omega_0 + Q(B, s)=0,
\eeq
where  $Q(B,s)$ is of order $g^2$, as before, and can again be
expanded in power series around $s=-i\omega_0$, in order to find
the pole perturbatively. However, this time one has to choose the
right determination for the function $Q(B,s)$. There are two cases:
a) The branch point $-iB$ is situated above $-i\omega_0$, so that
$-i\omega_0$ lies on both cuts. See Figure \ref{fig:tagli}(a); b)
The branch point $-iB$ is situated below $-i\omega_0$, so that
$-i\omega_0$ lies only on the upper branch cut. See Figure
\ref{fig:tagli}(b).
In case a), i.e.\ for $B<\omega_0$, the pole is on the third
Riemann sheet (under both cuts) and the power series converges in a
circle lying half on the first and half on the third Riemann sheet,
within a convergence radius $R_c = \omega_0-B$, which decreases as
$B$ increases [Figure~\ref{fig:tagli}(a)]. In case b), i.e.\ for
$B>\omega_0$, the pole is on the second Riemann sheet (under the
upper cut only) and the power series converges in a circle lying
half on the first and half on the second Riemann sheet, within a
convergence radius $R_c = B-\omega_0$, which increases with $B$
[Figure~\ref{fig:tagli}(b)]. In both cases we can write, for
$|s_{\rm pole}+i\omega_0|<R_c = |B-\omega_0|$,
\andy{poloB}
\barr
s_{\rm pole} &=&-i\omega_0-\frac{1}{2}\left\{Q[-i(\omega_0+B)+0^+]
+Q[-i(\omega_0-B)+0^+]\right\}+O(g^4) \nonumber \\
&=&-i \omega_0-\frac{1}{2}g^2 \omega_0\left\{q[-i(\omega_0+B)+0^+]
+q[-i(\omega_0-B)+0^+]\right\}+O(g^4) . \nonumber \\
\label{eq:poloB}
\earr
Our analysis is therefore valid when $|B-\omega_0|$ is larger than
$g^2 \omega_0$, (we remind the reader that we are interested in
large values of $B$). Equation (\ref{eq:poloB}) enables us to
analyze the behavior of the lifetime of level \#2.

\section{Decay rate vs $B$}

\begin{figure}
\centerline{\epsfig{file=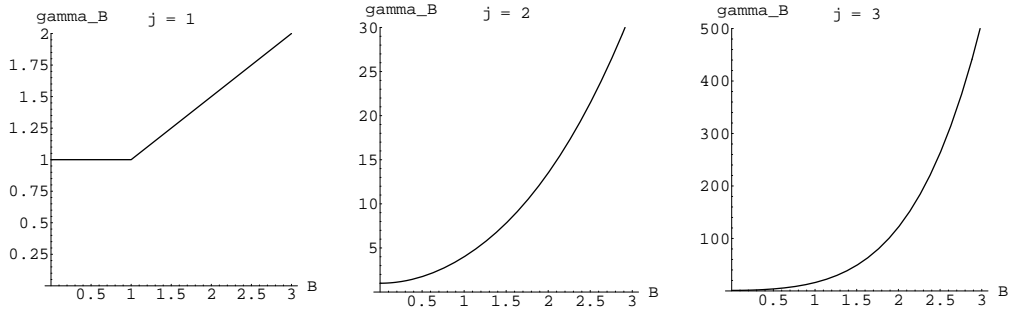,width=\textwidth}}
\caption{%
Fig.\ 4.1. The decay rate $\gamma(B)$ vs $B$, for electric
transitions with $j=1,2,3$; $\gamma(B)$ is in units $\gamma$ and
$B$ in units $\omega_0$. Notice the different scales on the
vertical axis.}
\label{fig:gamma(B)dip}
\end{figure}

We write, as in (\ref{eq:spole}),
$s_{\rm pole}=-i\omega_0+i\Delta E(B)-\gamma(B)/2$. By
substituting (\ref{eq:Q(-ieta)}) into (\ref{eq:poloB}) and taking
the real part, one obtains the following expression for the decay
rate
\andy{gamma(B)}
\beq\label{eq:gamma(B)}
\gamma(B)=\pi g^2 \omega_0 \left[\chi^2(\omega_0+B)+
\chi^2(\omega_0-B)\theta(\omega_0-B)\right] +O(g^4),
\eeq
which becomes, taking into account (\ref{eq:Fgr}),
\andy{MPSs}
\beq\label{eq:MPSs}
\gamma(B)= \frac{\gamma}{2} \; \frac{\chi^2(\omega_0+B)+
\chi^2(\omega_0-B)\theta(\omega_0-B)}{\chi^2(\omega_0)} +O(g^4).
\eeq
This is our main result and involves no approximations. It
expresses the lifetime $\gamma(B)^{-1}$, when the system is bathed
in an intense laser field $B$, in terms of the ``ordinary" lifetime
$\gamma^{-1}$, when there is no laser field. By taking into account
the general behavior (\ref{eq:chiprop}) of the matrix elements
$\chi^2(\omega)$ and substituting into (\ref{eq:MPSs}), one gets to
$O(g^4$)
\andy{gamma(B)dip}
\beq\label{eq:gamma(B)dip}
\gamma(B) \simeq
\frac{\gamma}{2}\left[\left(1+\frac{B}{\omega_0}\right)^{2j\mp 1}
+\left(1-\frac{B}{\omega_0}\right)^{2j\mp 1}
\theta(\omega_0-B)\right], \qquad (B \ll \Lambda)
\eeq
where $\mp$ refers to $1-2$ transitions of electric and magnetic
type, respectively. Observe that, since $\Lambda \sim$ inverse Bohr
radius, the case $B \ll \Lambda$ is the physically relevant one
\cite{FPinduced}. The decay rate is profoundly modified by the
presence of the ``$B$"-field. Its behavior is shown in
Figure~\ref{fig:gamma(B)dip} for a few values of $j$. The case
$j=1$ (transition of electric dipole type) yields a constant value
up to $B=\omega_0$; above this threshold, $\gamma$ increases
linearly with $B$. For $j > 1$ the derivative of $\gamma (B)$ is
continuous. In general, the decay rate $\gamma(B)$ increases with
$B$, so that the lifetime $\gamma(B)^{-1}$ decreases as $B$ is
increased. If one looks at $B$ as the strength of the
``observation" performed by the laser beam on level \#2 \cite{MPS},
one can view this phenomenon as an ``inverse" quantum Zeno effect.

Equation (\ref{eq:gamma(B)dip}) is valid for $B\ll\Lambda$. In the
opposite case, $B\gg\Lambda$, by (\ref{eq:chiprop}) and
(\ref{eq:MPSs}), one gets to $O(g^4$)
\andy{gammaom}
\beq\label{eq:gammaom}
\gamma(B) \sim \frac{\gamma}{2} \; \frac{\chi^2(B)}{\chi^2(\omega_0)}
\propto B^{-\beta}.
 \qquad (B \gg \Lambda)
\eeq
This is essentially the result obtained in Ref.\ \cite{MPS}. If
such high values of $B$ were experimentally obtainable, the decay
would be considerably hindered (quantum Zeno effect \cite{MPS}).

A thorough investigation of the phenomenon we have just proposed is
in preparation \cite{FPinduced}. It involves a complete discussion
in terms of Fano states and electromagnetic induced transparency
\cite{induced} and an analysis of the decay rates from level \#2 to
the dressed atomic or molecular states.

\medskip
\noindent {\bf Acknowledgments}
We thank E. Mihokova and L.S.\ Schulman for many comments and
useful discussions.



\begin{thebibliography}{99}

\bibitem{Dirac}\andy{Dirac}
P.A.M. Dirac, {\sl The principles of quantum mechanics} (Oxford
Univ. Press, London, 1958)

\bibitem{shortt} \andy{shortt}
A. Beskow and J. Nilsson, {\sl Arkiv f\"ur Fysik} {\bf 34} (1967)
561; L.A. Khalfin, {\sl Zh. Eksp. Teor. Fiz. Pis. Red.} {\bf 8}
(1968) 106 [{\sl JETP Letters} {\bf 8} (1968) 65]; L. Fonda, G.C.
Ghirardi, A. Rimini and T. Weber, {\sl Nuovo Cim.} {\bf A15} (1973)
689; {\bf A18} (1973) 805

\bibitem{longtt} \andy{longtt}
L. Mandelstam and I. Tamm, {\sl J. Phys.} {\bf 9} (1945) 249; V.
Fock and N. Krylov, {\sl J. Phys.} {\bf 11} (1947) 112; E.J.
Hellund, {\sl Phys. Rev.} {\bf 89} (1953) 919; M. Namiki and N.
Mugibayashi, {\sl Prog. Theor. Phys.} {\bf 10} (1953) 474. L.A.
Khalfin, {\sl Dokl. Acad. Nauk USSR} {\bf 115} (1957) 277 [{\sl
Sov. Phys. Dokl.} {\bf 2} (1957) 340]; {\sl Zh. Eksp. Teor. Fiz.}
{\bf 33} (1958) 1371 [{\sl Sov. Phys. JETP} {\bf 6} (1958) 1053]

\bibitem{seminal} \andy{seminal}
G. Gamow, {\sl Z. Phys.} {\bf 51} (1928) 204; V. Weisskopf and E.P.
Wigner, {\sl Z. Phys.} {\bf 63} (1930) 54; {\bf 65} (1930) 18; G.
Breit and E.P. Wigner, {\sl Phys. Rev.} {\bf 49} (1936) 519

\bibitem{Fermigold} \andy{Fermigold}
E. Fermi, {\sl Rev. Mod. Phys.} {\bf 4} (1932) 87; {\sl Nuclear
Physics} (Univ. Chicago, Chicago, 1950) pp.\ 136, 148; See also
{\sl Notes on Quantum Mechanics. A Course Given at the University
of Chicago in 1954} edited by E Segr\'e (Univ.\ Chicago, Chicago,
1960) Lec.\ 23

\bibitem{Sakurai}\andy{Sakurai}
J.J. Sakurai, {\sl Modern quantum mechanics} (Addison-Wesley,
Reading, Massachusetts, 1994); A. Messiah, {\sl Quantum mechanics}
(Interscience, New York, 1961)

\bibitem{Brown}\andy{Brown}
L.S. Brown, {\sl Quantum field theory} (Cambridge University Press,
Bristol, 1992)

\bibitem{temprevi}\andy{temprevi}
H. Nakazato, M. Namiki and S. Pascazio, {\sl Int. J. Mod. Phys.}
{\bf B10} (1996) 247

\bibitem{MPS} \andy{MPS}
E. Mihokova, S. Pascazio and L.S. Schulman, {\sl Phys. Rev.} {\bf
A56} (1997) 25

\bibitem{QZE} \andy{QZE}
B. Misra and E.C.G. Sudarshan, {\sl J. Math. Phys.} {\bf 18} (1977)
756; A. Peres, {\sl Am. J. Phys.} {\bf 48} (1980) 931; S. Pascazio,
M. Namiki, G. Badurek and H. Rauch, {\sl Phys. Lett.} {\bf A179}
(1993) 155; S. Pascazio and M. Namiki, {\sl Phys. Rev.} {\bf A50}
(1994) 4582; A. Beige and G. Hegerfeldt, Phys.\ Rev.\ {\bf A53}, 53
(1996) and references therein

\bibitem{induced}
\andy{induced}
U. Fano, {\sl Phys. Rev.} {\bf 124} (1961) 1866 ; K.J. Boller, A.
Imamoglu and S.E. Harris, {\sl Phys. Rev. Lett.} {\bf 66} (1991)
2593; J.E. Field, K.H. Hahn and S.E. Harris, {\sl Phys. Rev. Lett.}
{\bf 67} (1991) 3062; S.P. Tewari and G.S.Agarwal, {\sl Phys. Rev.
Lett.} {\bf 56} (1986) 1811; S.E. Harris, J.E. Field and A.
Imamoglu, {\sl Phys. Rev. Lett.} {\bf 64} (1990) 1811

\bibitem{Knight} \andy{Knight}
P.M. Radmore and P.L. Knight, {\sl J. Phys. (At. Mol. Phys.)} {\bf
B15} (1982) 561; P.L. Knight and M.A. Lauder, {\sl Phys. Rep.} {\bf
190} (1990) 1

\bibitem{TammDancoff} \andy{TammDancoff}
I. Tamm, {\sl J. Phys.} (USSR) {\bf 9} (1945) 449; S. Dancoff, {\sl
Phys. Rev.} {\bf 78} (1950) 382

\bibitem{BLP} \andy{BLP}
V.B. Berestetskii, E.M. Lifshits and L.P. Pitaevskii, {\sl Quantum
electrodynamics, Course of Theor. Phys., Vol. 4} (Pergamon Press,
Oxford, 1982), Chapter 5; H.E. Moses, {\sl Lett. Nuovo Cimento}
{\bf 4} (1972) 51; 54; {\sl Phys. Rev.} {\bf A8} (1973) 1710; J.
Seke, {\sl Physica} {\bf A203} (1994) 269; 284

\bibitem{FPinduced}\andy{FPinduced}
P. Facchi and S. Pascazio, in preparation

\bibitem{FP1} \andy{FP1}
P. Facchi and S. Pascazio, {\sl Phys. Lett.} {\bf A241}, 139 (1998)

\end{thebibliography}
\end{document}